\documentclass[10pt,letterpaper,nofonttune]{IEEEtran}
\usepackage{booktabs} 

\usepackage{epsfig}
\usepackage{amsmath}
\usepackage{subfigure}
\usepackage{graphicx}
\usepackage{cite}
\usepackage{amssymb}
\usepackage{fixmath}
\usepackage{caption}
\usepackage{algorithmic}
\usepackage{algorithm}
\usepackage[usenames]{color}
\usepackage[super]{nth}

\usepackage{algorithmic}
\usepackage{algorithm}
\usepackage{multirow}
\usepackage{comment}
\usepackage{flushend}
\usepackage{paralist}

\newdimen\figrasterwd
\figrasterwd\textwidth

\begin{document}
\title{\textit{CoBeam}: Beamforming-based Spectrum Sharing With Zero Cross-Technology Signaling for 5G Wireless Networks}




\author{\IEEEauthorblockN{
Lorenzo Bertizzolo$^\dagger$, 
Emrecan Demirors$^\dagger$,
Zhangyu Guan$^\ddagger$, 
Tommaso Melodia$^\dagger$\\
$^\ddagger$Institute for The Wireless Internet of Things,
Northeastern University, Boston, MA 02115, USA\\
$^\ddagger$Dept. of Electrical Engineering, The State University of New York (SUNY) at Buffalo, Buffalo, NY 14260, USA\\
Email: \{bertizzolo.l, demirors, melodia\}@northeastern.edu, guan@buffalo.edu \vspace{-10mm}}
\thanks{ 
This work was supported in part by the National Science Foundation under grant CNS-1618727.
%
}
\thanks{ 
This paper has been accepted for publication in IEEE INFOCOM 2020. This is a preprint version of the accepted paper. Copyright (c) 2020 IEEE. Personal use of this material is permitted. However, permission to use this material for any other purposes must be obtained from the IEEE by sending a request to pubs-permissions@ieee.org.
}
}

\clearpage\maketitle
\thispagestyle{empty}

\begin{abstract} This article studies an essential yet challenging problem in 5G wireless networks: \emph{Is it possible to enable spectrally-efficient spectrum sharing for heterogeneous wireless networks with different, possibly incompatible, spectrum access technologies on the same spectrum bands; without modifying the protocol stacks of existing wireless networks?} To answer this question, this article explores the system challenges that need to be addressed to enable a new spectrum sharing paradigm based on beamforming, which we refer to as {\em CoBeam}. In CoBeam, a secondary wireless network is allowed to access a spectrum band based on {\em cognitive beamforming} without mutual temporal exclusion, i.e., without interrupting the ongoing transmissions of coexisting wireless networks on the same bands; and without cross-technology communication. 
We first describe the main components of CoBeam, including \emph{programmable physical layer driver}, \emph{cognitive sensing engine}, and \emph{beamforming engine}, and then we showcase the potential of the CoBeam framework by designing a practical coexistence scheme between Wi-Fi and LTE on unlicensed bands. 
We present a prototype of the resulting coexisting Wi-Fi/U-LTE network built on off-the-shelf software radios based on which we evaluate the performance of CoBeam through an extensive experimental campaign.  
Performance evaluation results indicate that CoBeam can achieve on average $169\%$ throughput gain while requiring \emph{no} signaling exchange between the coexisting wireless networks. 
\end{abstract}


\section{Introduction}\label{sec:intro}
Mobile data traffic is increasing at an unprecedented rate, imposing a significant burden on the underlying wireless network infrastructure \cite{cisco2016global} - a problem often referred to as {\em spectrum crunch}. To address this challenge, researchers have been exploring new spectrum management/sharing approaches to increase the flexibility, spectral efficiency, and ultimately the network capacity of next-generation wireless networks (5G and beyond) \cite{Teng2017, bayhan2018future}.   
Along these lines, in this paper we study an essential yet challenging problem in 5G wireless networks. Specifically, we look at {\em how to enable spectrally-efficient spectrum sharing for heterogeneous wireless networks co-located in the same spectrum bands with different, possibly incompatible, spectrum access technologies.} 

Several challenges need to be addressed to enable efficient and practical spectrum sharing. First, the network throughput must be maximized while guaranteeing fairness among coexisting wireless systems that rely on different spectrum access techniques \cite{Qualcomm, cierny2017fairness}. 
Second, it is desirable for heterogeneous technologies to share the same spectrum and to achieve harmonious coexistence {\em without explicit cross-technology coordination signaling procedures}, which often require substantial modifications to the protocol stack. 
This becomes even more challenging when multiple uncoordinated service providers with diverging goals are involved  
\cite{li2010spectrum, khawer2016usicic, kosek2017coexistence}.

\begin{figure}[t]
\centering
\includegraphics[width=0.9\columnwidth]{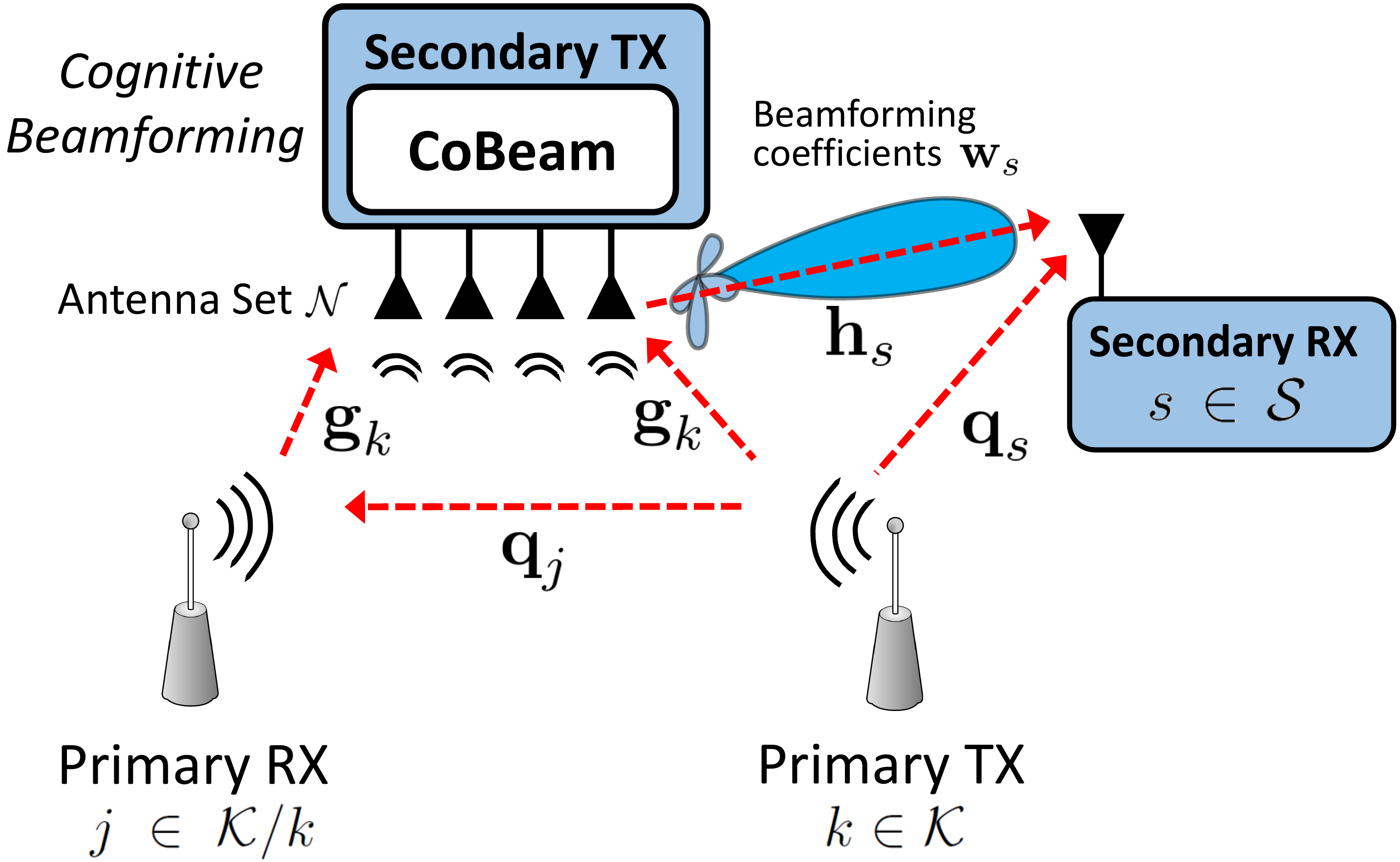} \caption{ \small  Beamforming-based spectrum sharing between primary and secondary technologies co-located on the same spectrum bands.}
\label{fig:scenario}
\vspace{-5mm}
\end{figure}

To address these challenges, significant  efforts have been put forward by industry and academia alike. For example, leading manufacturers, including Qualcomm, Ericsson, and Huawei, among others, have been collaborating with Service Providers (SPs) like Verizon and China Mobile to extend cellular wireless systems typically operating on licensed spectrum bands based on centrally scheduled channel access, e.g., LTE/LTE-A, 
to unlicensed bands so as to harvest additional spectrum resources~\cite{Qualcomm, LAA, Erika13}. 
To enable fair spectrum sharing with wireless systems that operate on unlicensed bands with carrier-sensing-based channel access, e.g., Wi-Fi, several coexistence mechanisms have been proposed, including LTE-U \cite{Qualcomm}, licensed assisted access (LAA) \cite{LAA}, enhanced LAA (eLAA) \cite{eLAA}, LTE-WLAN aggregation (LWA)~\cite{LWA} as well as MulteFire~\cite{MulteFire}. Readers are referred to \cite{Zinno2018, Bajracharya2018, Qualcomm000} and references therein for good surveys of this field. 

The common trait among these approaches is that they essentially base their spectrum sharing strategy on {\em mutual temporal exclusion}, i.e, they separate conflicting transmissions in the time domain.\footnote{We will discuss a few exceptions in Section~\ref{sec:relatedWorks}.} Therefore, \textit{at most one wireless system is allowed to access a given channel at each time instant} while other interfering systems must back off. 
In this paper, we take a different perspective. Our goal is to achieve higher spectrum utilization through a new spectrum sharing approach based on {\em cognitive beamforming}, which we refer to as CoBeam. 
CoBeam enables a multi-antenna secondary transmitter to access the channel without interfering with the primary system, by eavesdropping on the on-going primary user transmissions, and using the collected information to implement  beamforming-based transmissions (see Fig.~\ref{fig:scenario}) with minimal interference to the primary user.   
CoBeam leverages beamforming techniques and spatial diversity to enable multiple, co-located wireless technologies to access the same portion of the spectrum simultaneously without causing significant performance degradation to each other. Remarkably, CoBeam requires \emph{zero} signaling exchange among the coexisting technologies and \emph{no} modification to the protocol stack of previously-deployed systems. 
The main contributions of this article can be summarized as follows:
\begin{itemize}
\item \emph{CoBeam Framework Design.} 
We propose for the first time CoBeam, a new, cognitive-beamforming-based spectrum sharing approach for 5G-and-beyond wireless networks. We discuss the design of the main components of the CoBeam framework, including programmable physical layer driver, cognitive sensing engine, beamforming engine, and scheduling engine. 

\item \emph{Prototype Development.}  To demonstrate the effectiveness of the proposed framework, we present a prototype of CoBeam by considering a specific problem in 5G wireless networks, i.e., spectrum sharing between coexisting Wi-Fi and LTE in the same unlicensed spectrum bands.  

\item \emph{Experimental Performance Evaluation.} We extensively evaluate the performance of CoBeam on a large-scale office-space indoor testbed based on software-defined radios. Through extensive experiments, we show that an average of $169\%$  throughput gain can be achieved for the resulting coexisting Wi-Fi/U-LTE networks with guaranteed cross-technology fairness.

\end{itemize}

The remainder of the paper is organized as follows. 
We review related work in Section~\ref{sec:relatedWorks}, and formulate the spectrum sharing network problem in Section \ref{sec:problem}. 
In Section~\ref{sec:cobeam} we describe the CoBeam framework design architecture, and then discuss its prototyping by considering coexistence between Wi-Fi and U-LTE in the same unlicensed bands in Section~\ref{sec:wifilte}. Experimental performance evaluation is presented in Section~\ref{sec:evaluation}. Finally, we draw the main conclusions in Section~\ref{sec:conclusion}.

\section{Related Work} 
\label{sec:relatedWorks}

The topic of spectrum sharing between heterogeneous wireless systems has been the objective of intense research activities in recent years. For example, Guo et al. proposed a coexistence scheme between ZigBee and Wi-Fi networks based on adaptive forward error-correction coding \cite{PengGuo14}, while Chiasserini et al. studied in \cite{Chiasserini03} spectrum sharing between WLAN and Bluetooth systems in the $2.4~\mathrm{GHz}$ bands. A ``cooperative busy tone'' technique was proposed by Zhang et al. in \cite{xinyu11} to improve the visibility of ZigBee network to coexisting Wi-Fi networks. Spectrum sharing between Wi-Fi and LTE/LTE-A in the unlicensed bands has been studied in \cite{Erika13, Rapeepat12, Andre13, Timo13, Shweta14, Qiu2015, GuanINFOCOM16, bayhan2018future, abdelfattah2017modeling, YanHuang2018, Zhimin2017, gawlowicz2018enabling, Vlachou2018}.  For example, the listen-before-talk (LBT) scheme was introduced in \cite{Rapeepat12, makris2017measuring, wei2018joint} to enable carrier sensing at each LTE PeNB. It was shown in \cite{Erika13,Andre13} that improper assignment of almost-blank sub-frames (ABSF) for coexisting U-LTE networks can degrade the throughput of Wi-Fi networks significantly. Different from these pioneering contributions, where spectrum sharing is based on mutual temporal exclusion, we explore a new beamforming-based approach where co-located wireless networks are allowed to access the same channel simultaneously. 
Beamforming-based spectrum sharing has been studied in \cite{bayhan2017coexistence,  Zubow2018,  GeraciJSAC2017, Rodriguez2018}. For example, 
Geraci et al. present a MIMO-based interference rejection transmission scheme for cellular/Wi-Fi coexistence in \cite{GeraciJSAC2017, Rodriguez2018}. 
Bayhan et al. propose XZero in \cite{bayhan2017coexistence, Zubow2018}, a MIMO-based interference nulling framework for coexistence between LTE-U and Wi-Fi, relying on explicit cooperation via a cross-technology control channel.

Different from these approaches, which only provide simulation results without experimental evidence \cite{bayhan2017coexistence, Zubow2018, wei2018joint, GeraciJSAC2017, Rodriguez2018}, 
require signaling exchange between coexisting networks or modifications to the protocol stack of Wi-Fi networks \cite{Zubow2018, bayhan2017coexistence}, 
CoBeam is a new cognitive-beamforming-based spectrum sharing approach that requires \emph{zero} cross-technology signaling and \emph{no} changes to protocol stacks of existing wireless networks.

\section{Spectrum Sharing Network Problem Formulation } \label{sec:problem}
In this section, we first formalize the spectrum sharing design objective and challenges. Then, we provide a detailed description of the CoBeam framework and its main components in Section \ref{sec:cobeam}. 
For the sake of convenience, in the following description, we refer to an incumbent (possibly licensed) wireless network as \emph{primary system}, and to a second wireless network intended to coexist with the primary system as \emph{secondary system}. 
We herein investigate the spectrum efficiency goals and the challenges associated with heterogeneous 5G wireless networks where a primary and a secondary system coexist on the same spectrum bands (see Fig.~\ref{fig:scenario}). 
Specifically, the primary system consists of a set $\mathcal{K}$ of single-antenna transceivers communicating with each other in a point-to-point, multicast, or broadcast fashion. 
As shown in Fig.~\ref{fig:scenario} the secondary system, instead, is a $\mathcal{N}$-antenna transceiver willing to communicate with a set $\mathcal{S}$ of single-antenna secondary receivers. We let $\mathcal{N}$ denote the set of antennas available at the secondary transmitters.

We first introduce some notations. We divide the transmission time into a set $\mathcal{T}$ of consecutive time slots. Block fading channel is considered, i.e., the wireless channel is considered to be fixed in each time slot \mbox{$\nu\in\mathcal{T}$}. 
Let us define the channel coefficients between the primary and secondary systems as illustrated in Fig.~\ref{fig:scenario}.
Let $\textbf{h}_s^\nu = \left(\sqrt{\bar{h}_{ns}^\nu}\tilde{h}_{ns}^\nu\right)_{n\in\mathcal{N}}$ 
denote the channel gain vector between secondary transmit antennas in $\mathcal{N}$ and secondary receiver $s\in\mathcal{S}$, in time slot $\nu\in\mathcal{T}$, that is, the channel of the secondary signal, where $\bar{h}_{ns}^\nu$ and $\tilde{h}_{ns}^\nu$ represent the path loss and the small-scale fading coefficient, respectively. 
Similarly, let 
$\textbf{g}_k^\nu = \left(\sqrt{\bar{g}_{nk}^\nu}\tilde{g}_{nk}^\nu\right)_{n\in\mathcal{N}}$ 
denote the channel coefficient vector between the secondary transmit antennas in $\mathcal{N}$ and primary user $k\in\mathcal{K}$ in time slot $\nu\in\mathcal{T}$, 
that is, the channel causing interference to the primary system.
Finally, let $\textbf{q}_{j}^\nu = \left(\sqrt{\bar{q}_{kj}^\nu}\tilde{q}_{kj}^\nu \right)_{k\in\mathcal{K}}$ 
and $\textbf{q}_{s}^\nu = \left(\sqrt{\bar{q}_{ks}^\nu}\tilde{q}_{ks}^\nu \right)_{k\in\mathcal{K}}$ 
denote the vector of channel coefficients between a  primary transmitter $k \in \mathcal{K}$ and a primary receiver $j\in \mathcal{K}/k$, and a secondary receiver $ s \in\mathcal{S} $, respectively. 
As shown in Fig.~\ref{fig:scenario}, the two channel vectors $\textbf{q}_{j}^\nu $ and $\textbf{q}_{s}^\nu $ represent the channel of the signal to the primary system and the channel of the received interference to the secondary system, respectively.

Referring to a scenario as in Fig.~\ref{fig:scenario} where a multi-antenna secondary transmitter is willing to communicate with a set of single-antenna secondary receivers, let us represent the secondary transmitted signal in time slot 
$\nu\in\mathcal{T}$ 
as 
\begin{equation}
\label{eq:transmitter_secondary}
\textbf{x}^\nu = \sum_{s\in\mathcal{S}_{\mathrm{sch}}}\sqrt{p_{s}}\textbf{w}_{s}^\nu x_{s}^\nu,
\end{equation}
\noindent where 
$\mathcal{S}_{\mathrm{sch}} \subset \mathcal{S}$ 
is the scheduled set of secondary users to be served in time slot $\nu$.
In the above formulation \eqref{eq:transmitter_secondary}, 
$p_{s}$
is the secondary transmitter signal power,
$x_s^\nu$ is the data symbol intended for the secondary receiver $s\in\mathcal{S}_{\mathrm{sch}}$,  and $\textbf{w}_{s}^\nu = (w_{ns}^\nu)_{n\in\mathcal{N}}$ 
is the beamforming vector employed at the $\mathcal{N}$-antenna secondary transmitter, where $w_{ns}^\nu$ is the channel coefficient between antenna $n\in\mathcal{N}$ and secondary receiver $s$. 

Consequently, according to the channel gain vectors formulations above, the corresponding signal-to-interference-plus-noise ratio (SINR) at primary $\mathcal{K}$ and secondary  ${S}_{\mathrm{sch}}$ receivers in time slot $\nu\in\mathcal{T}$ can be expressed as

\small
\begin{equation}
\label{formula:sinr_primary}
\text{SINR}_{s}^{\nu} = 
\frac
{|\sqrt{p_{s}}\textbf{w}_{s}^{\nu}\textbf{h}_{s}^{\nu}
|^2}
{\bigg|
\sum\limits_{s' \in \mathcal{S_\mathrm{sch}}/s}  \sqrt{p_{s'}}\textbf{w}_{s'}^{\nu}\textbf{h}_{s'}^{\nu}
\bigg|^2
+
{\left|
\sum\limits_{k \in \mathcal{K}}   \sqrt{p_{k}}q_{ks}^{\nu}
\right|^2
}
+ \sigma_{s}^{2}
}
\end{equation}
\normalsize

\small
\begin{align}
\label{formula:sinr_secondary}
\text{SINR}_{j}^{\nu} = 
\frac
{\left|
  \sqrt{p_{k}}q_{kj}^{\nu} 
\right|^2
}
{\bigg|
\sum\limits_{s \in \mathcal{S_\mathrm{sch}}}  \sqrt{p_{s}}\textbf{w}_{s}^{\nu} g_{k}^{\nu}
\bigg|^2
+
{\bigg|
\sum\limits_{
\substack{k' \in \mathcal{K} /k \\ j' \in \mathcal{K} / j}
}   \sqrt{p_{k'}}q_{k'j}^{\nu}
\bigg|^2
}
+ \sigma_{k}^{2}
}
\end{align}
\normalsize
for secondary receiver $s\in\mathcal{S}_{\mathrm{sch}}$ and
for primary receiver $j\in \mathcal{K}/k$ when served by primary transmitter $k\in \mathcal{K}$, respectively. 

Based on the above problem formulation, a harmonious spectrum coexistence scheme is achieved when the secondary transmitter beamforming vectors  
$ \{ \textbf{w}_s^\nu \}, s \in \mathcal{S}_{\mathrm{sch}}$
guarantee satisfactory SINR levels at the secondary receivers, while not degrading the SINR at the primary receivers.
The objective of CoBeam is to determine, for each time slot~$\nu$, the optimal beamforming vectors $\{ \textbf{w}_s^\nu \}$ maximizing the spectrum utilization, 
i.e., $ {\arg\max} \{ (\text{SINR}_{s}^{\nu})_ {s \in \mathcal{S}_{\mathrm{sch}}} + (\text{SINR}_{k}^{\nu})_ {k\in\mathcal{K}} \}$,
while guaranteeing high fairness values between the coexisting wireless networks, e.g.,
$ {\arg\max} \{  
(\sum_{s \in \mathcal{S}_{\mathrm{sch}}  }^{k\in\mathcal{K}} \text{SINR}_{s}^{\nu}+\text{SINR}_{k}^{\nu}
)^2
/
[ (|\mathcal{K}| + |\mathcal{S}_{\mathrm{sch}|} )
*
\sum_{s \in \mathcal{S}_{\mathrm{sch}}}^{k\in\mathcal{K}} (\text{SINR}_{s}^{\nu})^2+(\text{SINR}_{k}^{\nu})^2
]
\}
$.

The spectrum sharing optimization problem discussed above is however not easy to solve in practical settings, as it faces some fundamental challenges that we summarize as follows:
\begin{itemize}
\item \textit{Lack of coordination among coexisting networks.} 
Low-interference spectrum sharing in cross-technology coexistence scenarios is often achieved via coordinated channel access, which requires signaling exchange among the coexisting wireless technologies. In turn, this may require significant modifications to the primary system's standard protocol stack. Conversely, here we propose to achieve spectrally-efficient and fair spectrum sharing with \emph{zero} cross-technology signaling exchanges.

\item \textit{Lack of a central controller.}
Optimal sharing of network resources   (e.g., bands, spectrum access) can be achieved through a central controller with a holistic view of the spectrum access strategies at the different network nodes. However, this requires significant signaling exchange, possibly dedicated infrastructure, and results in a single point of failure. Instead, 
CoBeam seeks to achieve optimal spectrum sharing without a central controller but in a fully distributed fashion.

\item \textit{Backward compatibility and transparency:} 
The spectrum sharing mechanisms mentioned above (i.e., time coordination, signaling, central controller, etc.) often require modifications to the upper layers of the protocol stack \cite{Vlachou2018, Zubow2018, bayhan2017coexistence}. 
Differently, CoBeam works at the very low layers of the protocol stack and requires no modification to upper layer protocols. 
To this end, CoBeam is backward compatible and it can be integrated into the secondary transmitter protocol stack without requiring any upper layer adaptation. Furthermore, once installed at the secondary system, {\em CoBeam is completely transparent to the primary system and the secondary receivers}. 

\end{itemize}
\noindent The next section will describe in detail CoBeam, a new cognitive-beamforming-based framework for secondary transmitters. We will first provide an overview of the three-module architecture design of CoBeam and then discuss in detail the functionalities implemented in each of its modules, namely \emph{Programmable Physical Layer Driver}, \emph{Cognitive Sensing Engine}, and \emph{Beamforming Engine}.

\section{CoBeam Framework Architecture Design}
\label{sec:cobeam}
\begin{figure}[t]
\centering
\includegraphics[width=0.9\columnwidth]{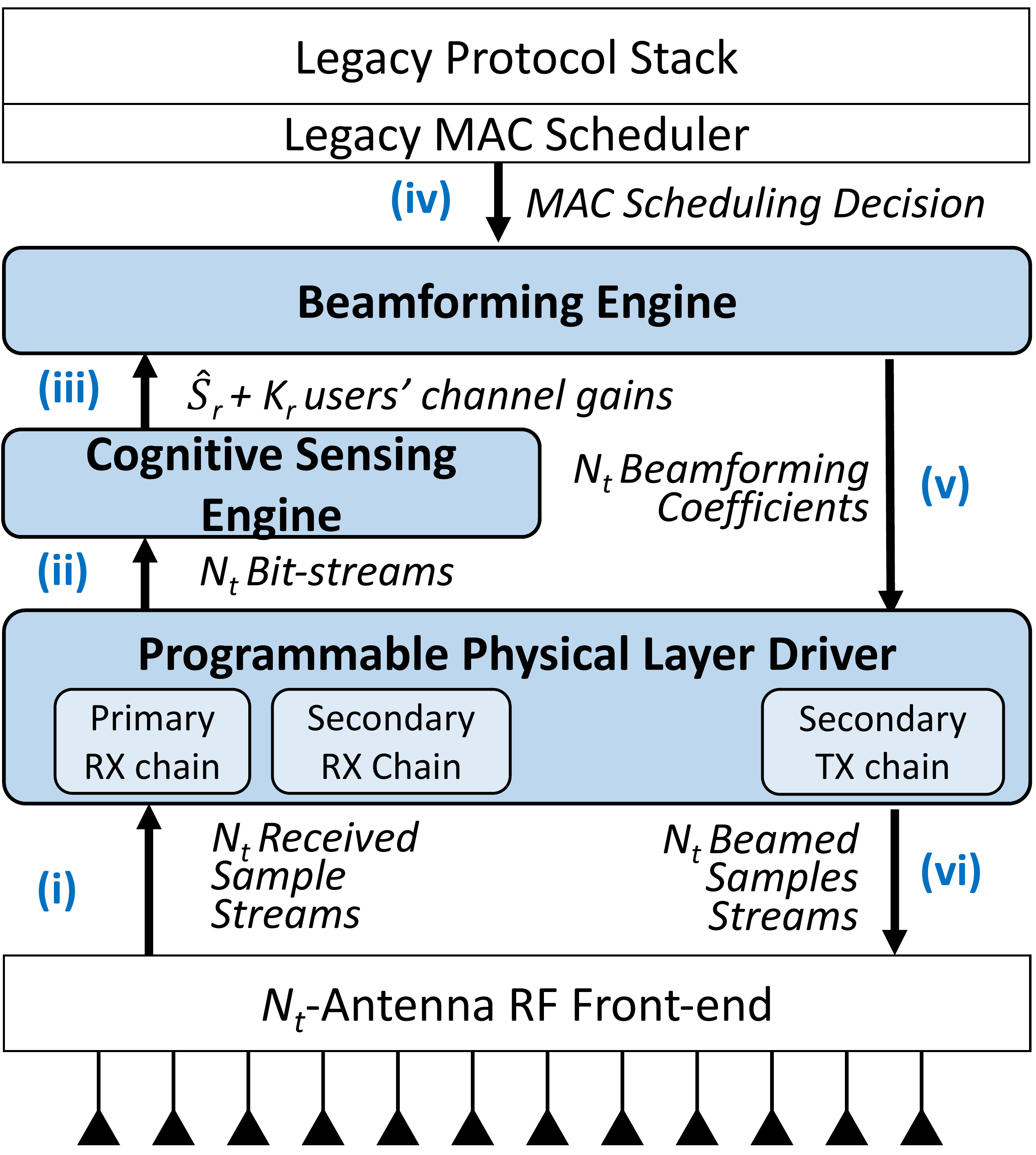} \caption{ \small  CoBeam architecture design overview. }
\label{fig:CoBeamframework}
\vspace{-5mm}
\end{figure}

As illustrated in Fig.~\ref{fig:CoBeamframework}, the CoBeam architecture is structured into three distinct yet tightly interacting components, namely \textit{Programmable Physical Layer Driver}, \textit{Cognitive Sensing Engine}, and \textit{Beamforming Engine}. 
The Programmable Physical Layer Driver interfaces the transmitter and receiver chains of the radio front-end, and interfaces with the Beamforming Engine through the Cognitive Sensing Engine. On the other hand, the Beamforming Engine interacts with the legacy Medium Access Control (MAC) scheduler of the secondary transmitter. In this way, Cobeam operates at the bottom layer of the protocol stack, below the MAC. 

Referring to Fig. \ref{fig:CoBeamframework},
the \textit{Programmable Physical Layer Driver} interfaces the transmitter and receiver chains of the RF front-end (i), demodulates the received samples, and passes them to the \textit{Cognitive Sensing Engine} (ii), which performs a wireless channel analysis targeted to extract the channel gains. The latter is passed to the \textit{Beamforming Engine} (iii), which interacts with the legacy Medium Access Control (MAC) scheduler of the secondary transmitter to calculate the beamforming coefficients for the secondary users to be served  (iv). Lastly, these coefficients are passed at the \textit{Programmable Physical Layer Driver }(v), which precodes and modulates the data bit-streams to be sent to the RF front-end (vi).

\noindent $\bullet$
\textbf{Programmable Physical Layer Driver.} 
For each available antenna, the \textit{Programmable Physical Layer Driver} implements the secondary system transmission and receiver chains, which can be in multiple wireless technologies, e.g., Wi-Fi, LTE, ZigBee, and Bluetooth.
Moreover, this module implements an additional receiver chain able to receive and demodulate primary users data packets (e.g., Wi-Fi), for a total of two receiver chains and one transmitter chain for each available antenna.
In the receiver chains, the driver demodulates baseband digital samples into bit-streams that are passed up to the \textit{Cognitive Sensing Engine} for traffic analysis. 
In the transmitter chains, the Driver performs physical layer spectrum access, employing the beamforming coefficients calculated by the \textit{Beamforming Engine} to precode and then modulate the data to be transmitted. 
This module operates at the physical layer of the protocol stack to demodulate wireless signals coming from the radio front-end into bit-streams, and vice-versa.
The rationale behind its operations is to hide the physical-layer details of the diverse coexisting wireless technologies to the upper layers of the protocol stack, and hence ensure CoBeam's transparency and preserve its backward compatibility. 

\noindent $\bullet$
\textbf{Cognitive Sensing Engine.} 
Based on the demodulated bit-streams fed by the \textit{Programmable Physical Layer Driver}, this module's primary task is to extract channel gain information and perform primary users' traffic analysis.

The channel gain information, e.g., CSI $\textbf{h}_s^\nu$, $\textbf{g}_k^\nu$ and $\textbf{q}_s^\nu$ for time slot $\nu$ defined in Section~\ref{sec:problem}, represents the channel characteristics between a primary or secondary users and the secondary transmitter where CoBeam is operating. 
This information, which accounts for the effects of distance, path loss, small and fast fading on the wireless channels, can be estimated based on a-priori knowledge of the transmitted signals such as known pilot symbols, golden sequences, or preambles. 
An example of channel gain estimation will be discussed later in Section~\ref{sec:wifilte}.   
It is worth pointing out that it is practical for the secondary transmitter to perform the channel estimation if it has the a-priori knowledge of the transmitted signals, e.g., when the primary system adopts a known or standard-defined physical layer preamble;
otherwise, blind or semi-blind channel estimation techniques can be employed by the secondary systems \cite{papadias1997space}. 

Through the channel gain estimation, the \textit{Cognitive Sensing Engine} is also able to detect the presence of the ongoing traffic of the primary system and further analyze its traffic pattern.
Note that all the operations in the \textit{Cognitive Sensing Engine} are performed upon demodulated data coming from the \textit{Programmable Physical Layer Driver}, and does not require any coordination with the primary system users, that is, no cross-technology signaling is needed. This ensures CoBeam's transparency to primary users.
The estimated channel gains are the necessary building blocks of any beamforming-based precoding techniques, and they are, once extracted, passed to the \textit{Beamforming Engine}. 

\begin{figure*}[t]
\centering
\includegraphics[width=1\textwidth]{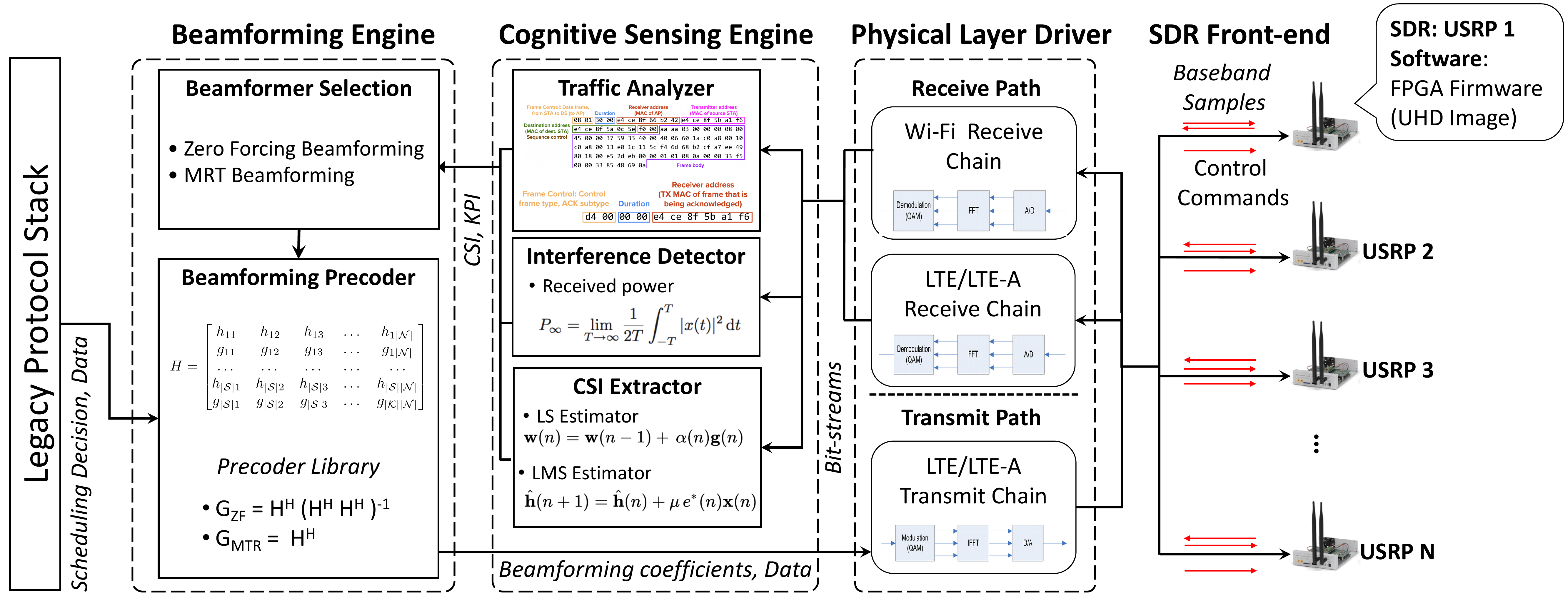} \caption{ \small  CoBeam Prototyping Diagram. }
\label{fig:prototype}
\vspace{-5mm}
\end{figure*}

\noindent $\bullet$
\textbf{Beamforming Engine.} 
Apart from the channel gain information, the \textit{Beamforming Engine} is also fed with a primary user traffic activity Key Performance Indicator (KPI), based on which the \textit{Beamforming Engine} selects the optimal spectrum sharing beamforming technique and then calculates the corresponding beamforming coefficients (i.e., $\mathbf{W}_s$ in Fig.~\ref{fig:scenario}). 

The KPI represents whether there is ongoing primary system traffic, the impact that the primary system has on the secondary transmitter at which CoBeam is installed, as well as, based on channel reciprocity of TDD systems, the impact of interference from omnidirectional secondary transmissions on the primary users.
Basing its decision upon the KPI, the \textit{Beamforming Engine} selects the optimal beamforming scheme that maximizes the spectrum utilization while guaranteeing fairness between the coexisting wireless networks. 
Optimal choices can be, for example, Maximum Ratio Transmission (MRT) \cite{MRT}, which maximizes the spectrum utilization favoring the secondary system in case of non-detected primary technology;
or Zero-Forcing (ZF) beamforming \cite{zfbeam}, which can be employed to deliver data at secondary receivers minimizing the interference caused at the primary system by nulling the received power at primary users in case of intense primary user channel activity.

Upon determining the optimal beamforming scheme, the beamforming engine constructs the optimal beamforming coefficient vectors based on the channel gains passed by the Cognitive Sensing Engine and on the secondary user schedule selection performed by the legacy Medium Access Control (MAC) protocol stack scheduler.
Specifically, given the set of scheduled secondary users $\mathcal{S}_{\mathrm{sch}}$ to be served in time slot $\nu \in \mathcal{T}$, the number of detected primary users sharing the spectrum $\mathcal{K}$, and the number of antennas at the radio front-end $\mathcal{N}$, the Beamforming Engine calculates the optimal beamforming coefficients matrix $\textbf{G}$:
\begin{equation}
\begin{aligned}
   \textbf{G} \in \mathbb{C}^{N_t \times ( \hat{S}_r + K_r )} = \{ \textbf{w}_{s} \}, \quad
   s \in \{ 1, \dots, \hat{S}_r + K_r \},
   \end{aligned}
\end{equation}
where $\textbf{w}_{s} =  \{ w_{ns} \}\;, n \in \{ 1, \dots, N_t \}  $ is the set of precoding coefficients for the wireless link from the $n$-th transmitting antenna to primary or secondary user $s$, 
and $N_t = |\mathcal{N}|$, $\hat{S}_r=|\mathcal{S}_{\mathrm{sch}}|$ and $K_r=|\mathcal{K}|$ represent the cardinality of the sets described above.
Depending on the selected beamforming scheme, these coefficients are passed back down to the Physical Layer Driver, which employs them to precode the users' data bit-streams in each of the $\mathcal{N}$ secondary system transmitter chains, prior to modulation. 

The result of CoBeam's lower-layer operations is a cognitive-based beamforming module that attempts to achieve fair spectrum sharing coexistence among heterogeneous technologies, and that can be seamlessly integrated at any secondary system protocol stack without any protocol modification or cross-technology signaling required.
In the following section, we illustrate in detail a prototype of CoBeam for LTE secondary operations in the unlicensed spectrum, designed to guarantee coexistence with Wi-Fi.

\section{CoBeam-based Spectrum Sharing for LTE in Unlicensed Bands} \label{sec:wifilte}
In this section, we discuss in detail the implementation of CoBeam in a multi-antenna LTE secondary transmitter in the unlicensed industrial, scientific and medical (ISM) bands populated by Wi-Fi devices, among others. 
Specifically, we prototyped CoBeam over a testbed based on Universal Software Radio Peripheral (USRP) N210s \cite{Ettus}, a commercial software-defined radio (SDR) front-end, to realize the coexistence of Wi-Fi and LTE/LTE-A networks on the same unlicensed bands. The prototyping diagram is illustrated in Fig.~\ref{fig:prototype}. 
We herein provide the implementation details of the three components described in Section \ref{sec:cobeam}, and then present the spectrum sharing results obtained through an extensive experimental campaign in Section \ref{sec:evaluation}.
In the following, we refer to LTE in Unlicensed band as U-LTE.

\noindent
$\bullet$ \textit{Programmable Physical Layer Driver.}
In this application scenario, CoBeam incorporates two receiver chains in the physical layer module. 
The first is based on IEEE 802.11 to eavesdrop on the Wi-Fi traffic, while the second receives traffic transmitted by U-LTE user terminals.  
The Wi-Fi receiver is based on the Orthogonal-Frequency-Division-Multiplexing (OFDM), which leverages a number of closely spaced orthogonal subcarriers. Specifically, it adopts the 802.11a packet format that consists of a preamble and a payload. 
The preamble contains two standard-defined training fields that are used for time and frequency synchronization and channel state information (CSI) estimation. 
Each Wi-Fi OFDM symbol contains $52$ subcarriers, where $48$ of them are used for data symbols while the rest are pilot symbols for tracking frequency, phase, and amplitude variations over the burst. 
The LTE receiver chain is instead based on  Single Carrier Frequency Division Multiple Access (SC-FDMA) as specified in the LTE uplink standard \cite{holma2009lte}. SC-FDMA uses closely spaced orthogonal subcarriers similar to the OFDM and OFDMA schemes with a special precoding process to minimize the peak-to-average-power ratio (PAPR) and accordingly optimize the power consumption.

On the transmitter side, CoBeam implements an LTE downlink, which is based on an Orthogonal-Frequency-Division Multiple Access (OFDMA) scheme, which defines a physical resource block as the smallest unit of resources that can be allocated to a user. It contains $12$ adjacent subcarriers over a time slot (i.e., $0.5\:\mathrm{ms}$). While each subcarrier has a bandwidth of $15\:\mathrm{kHz}$, the system bandwidth can vary between $1.4\:\mathrm{MHz}$ and $20\:\mathrm{MHz}$. 
The transmitter chain leverages a  beamforming precoder which receives the precoding coefficients $\textbf{G}$ from the Beamforming Engine. In the transmitter chain, the received upper layer bit-streams are digitally precoded with the received coefficients, modulated, and then fed to the multi-antenna radio front end.

\noindent $\bullet$
\textit {Cognitive Sensing Engine.} 
The Cognitive Sensing Engine performs wireless channel gain extraction and primary user traffic pattern analysis to provide necessary physical- and MAC-layer information for cognitive beamforming. 
Specifically, it incorporates three main sub-modules, namely, \emph{Interference Detector, CSI Extractor,} and \emph{Traffic Analyzer}, to process the Wi-Fi and LTE bit-streams collected through the physical layer module discussed in Section~\ref{sec:cobeam}. 
All sub-modules operate simultaneously and synchronously minimizing processing delays and generate consistent CSI estimates that match the fast varying channel characteristics.
The functionalities of each sub-module are described as follows.

(i) Based on a-priori information on the receiver and transmitter OFDM parameters including center frequency, bandwidth, occupied tones and cyclic prefix length, the \emph{interference detector} evaluates the power of the received interference and noise at each of the $N_t$ U-LTE antennas. 
This is accomplished by integrating the intensity of the received Wi-Fi symbols every time an 802.11a preamble is detected, accounting for the frequency of received Wi-Fi packets. 
(ii) Standard-defined physical-layer 802.11a synchronization sequences and known secondary users LTE/LTE-A preambles also trigger the \emph{CSI extractor}. This module employs different estimators, such as Least Square (LS) and Least Mean Square (LMS) to extract CSI information through analyzing pilot OFDM symbols and pilots populated subcarriers to track down channel attenuation and phase rotation effects. 
The CSI is extracted then associated with the corresponding frame every time a U-LTE or Wi-Fi synchronization sequence is detected, for all the  $\mathcal{N}$ secondary transmitter antennas. 

(iii) Merging the Wi-Fi traffic traces with the Wi-Fi CSI obtained from the CSI extractor and the average and received power evaluated at the interference detector, the \emph{traffic analyzer} performs primary user traffic analysis.
It calculates a running average estimation of the primary user traffic and comes up with a KPI expressing the intensiveness of the surrounding Wi-Fi activity, for example, negligible, or pervasive. 
This information is fed to the Beamforming Engine.

\noindent $\bullet$
\textit{Beamforming Engine.} 
As presented in Section \ref{sec:cobeam}, the goal of this module is two-fold: 
i) determine, based on primary system activity KPI received from the \emph{traffic analyzer} the best beamforming scheme to maximize the spectrum efficiency while ensuring fair coexistence between the Wi-Fi and U-LTE networks; and 
ii) calculate the precoding coefficients for the selected beamforming scheme starting from the most recent CSI information for the scheduled set of users, in every time slot.
Note that because of the possible inaccuracy of the CSI estimates, the effectiveness of a precoding scheme is dictated by the degrees of freedom expressed by the ratio between the number of available antennas and the number of users whose channels are involved building the precoder $N_t / \big(S_s + K_r \big) > 1$, $K_r  = |\mathcal{K}|$, and $S_s = |\mathcal{S_{\mathrm{sch}}}|$. 

In the prototyped framework, the optimal beamforming scheme selection relies upon a threshold-based binary decision involving the received Wi-Fi activity indicator.
Such selection is driven by spectrum efficiency considerations, i.e.,  it adaptively favors beamforming schemes that attempt to achieve higher signal gain, while preserving the fairness between the coexisting Wi-Fi and U-LTE networks.

Specifically, in the prototyped framework, Maximum Ratio Transmission (MRT)~\cite{MRT} Beamforming is preferred in scenarios with low Wi-Fi traffic loads. 
Denoting $\textbf{H}_{mrt}$ as the channel matrix from the $\mathcal{N}$ secondary U-LTE transmitter antennas to the set of scheduled $\mathcal{S_{\mathrm{sch}}}$ secondary receivers, and being $(u_s)_{s\in\mathcal{S_{\mathrm{sch}}}} $ the vector of intended data for the users $ s \in \mathcal{S_{\mathrm{sch}}}$, the MRT beamforming coefficients are based on the following linear precoder 
\begin{equation}
\textbf{G} \in \mathbb{C}^{N_t \times S_s} = \textbf{H}^H_{mrt}.
\label{eq:mrt}
\end{equation}
\noindent
Consequently, the signal received at the $ s \in \mathcal{S_{\mathrm{sch}}}$ U-LTE users, denoted as $\textbf{y}$, is given by 
\begin{equation}
\begin{split}
    \textbf{y} \hspace{-0mm}  = \hspace{-0mm} 
    \begin{bmatrix}
     y_1 \\ 
     y_2 \\ 
     \dots \\
     y_{S_s}
     \end{bmatrix}
    \hspace{-0mm}  =    
     \textbf{G}\hspace{-0mm}  \times \textbf{H}_{mrt} \times \hspace{-0mm} 
    &\begin{bmatrix}
     u_1\\ 
     u_2 \\ 
     \dots \\
     u_{S_s}
     \end{bmatrix}
   =  \hspace{-0mm} 
| \textbf{H}_{mrt} |^2
\hspace{-0mm} \times \hspace{-0mm} 
    \begin{bmatrix}
     u_1 \\ 
     u_2 \\ 
     \dots \\
     u_{S_s}
     \end{bmatrix}
     \end{split}.
     \label{eq:mrt_ffective_channel}
\end{equation}
This scheme results in a distortion-free effective channel, maximizing the signal power gain at the intended U-LTE users.

Differently, in high Wi-Fi traffic load scenarios,
conservative low-interference beamforming schemes are preferred in favor of better network fairness, such as Zero-Forcing  (ZF) beamforming \cite{zfbeam}. 
Denoting $\textbf{H}_{zf}$ as the compound channel matrix from the secondary U-LTE transmitter antennas in $\mathcal{N}$ to the scheduled secondary U-LTE receivers in $\mathcal{S_{\mathrm{sch}}}$ and the eavesdropped Wi-Fi users in $\mathcal{K}$, the ZF beamforming scheme is obtained based on the precoder \textbf{G} defined as
\begin{equation}
\textbf{G} = \in \mathbb{C}^{N_t \times ( S_s + K_r )}  = \textbf{H}_{zf} \times \big(  \textbf{H}_{zf}^{H} \times \textbf{H}_{zf} \big)^{-1}.
\label{eq:zf}
\end{equation}
As a consequence of employing \eqref{eq:zf} as a precoder the resulting signal received at each network user $\textbf{y}$ is given by
\begin{align}
\begin{split}
    \textbf{y} \hspace{-0.8mm} = \hspace{-0.8mm}
    \begin{bmatrix}
     y_1 \\ 
     y_2 \\ 
     \dots \\
     y_{S_s + K_r}
     \end{bmatrix}
    \hspace{-0.8mm} = 
\textbf{G}\hspace{-0.8mm}  \times \textbf{H}_{zf} \times 
    &\begin{bmatrix}
     u_1 \\ 
     u_2 \\ 
     \dots \\
     u_{S_s + K_r}
     \end{bmatrix}
\hspace{-0.8mm} = 
\textbf{I} \hspace{-0.8mm}
\times \hspace{-0.8mm}
    \begin{bmatrix}
     u_1 \\ 
     u_2 \\ 
     \dots \\
     u_{S_s + K_r}
     \end{bmatrix}
     \end{split}.
     \label{eq:zf_effective_channel}
\end{align}
\normalsize
This beamforming scheme results in a diagonal effective channel minimizing the cross-user interference among U-LTE users while nulling the received signal power at the primary Wi-Fi network users.
Then, the signal-to-interference-plus-noise ratio (SINR) in (\ref{formula:sinr_primary}) can be rewritten as
   \begin{equation}
         \text{SINR}_{{mrt}} = \frac{P_s N_t}{{S_s} \big(  P_s + 1 \big) }
 \label{eq:snr_mrt}        
 \end{equation}
and
 \begin{equation}     \text{SINR}_{{zf}} = P_s \times  \frac{N_t - (S_s + K_r)}{{S_s + K_r}} \quad 
 \label{eq:snr_zf}        
 \end{equation}
 for MRT precoding and ZF precoding, respectively, where $P_s$ is the total transmission power employed at the secondary U-LTE transmitter \cite{zfmrt}.

At every time slot, the Beamforming Engine calculates the precoding coefficients for the scheduled set of secondary U-LTE users to serve, starting from the per-user CSI as in (\ref{eq:mrt}) and (\ref{eq:zf}).
The resulting precoding coefficients are finally passed down to the Physical Layer Driver and employed in the transmit chain to weigh the physical-layer bit-streams before modulating and feeding them to the SDR front-end for transmission.

\section{Performance Evaluation}\label{sec:evaluation}
In this section, we present extensive experimental results obtained implementing CoBeam on a Software-Defined Radios (SDRs)-based testbed employing USRP N210 SDR devices.

\subsection{Experimental Setup}
The testbed consists of a U-LTE network co-located with a series of Wi-Fi users on the same $2.4\;\mathrm{GHz}$ spectrum band. 
In all of the performed experiments, a secondary U-LTE transmitter endowed with up to four antennas transmits data toward one single-antenna secondary U-LTE receiver, while a primary system transmission between a Wi-Fi transmitter (AP) and a Wi-Fi receiver (STA) is ongoing at the same time.  
Both the U-LTE and Wi-Fi networks have been prototyped on USRP N210 based on a GNU Radio implementation \cite{Blossom04}. 
The Wi-Fi nodes are based on an IEEE 802.11a standard-compliant implementation~\cite{Bloessl13SRIF}.
IEEE 802.11 Channel 13 is used in the experiments with center frequency of $2.472\;\mathrm{GHz}$. The bandwidth is set to $5\;\mathrm{MHz}$ for the Wi-Fi network and $1\;\mathrm{MHz}$ for U-LTE, corresponding to approximately five LTE standard resource blocks. 
It is worth mentioning that the CoBeam prototype that we present here operates over narrower bandwidths than the ones defined in LTE and Wi-Fi standards because of the resource-hungry host-based baseband processing typical of SDRs implementations, which would result in unacceptable delays otherwise.
However, in CoBeam, the operational bandwidth is a programmable physical layer parameter. More powerful host computing machines, or parallel base-band processing would allow CoBeam to operate over standard-compliant bandwidths. 

First, we provide a simple example of spectrum sharing where U-LTE and Wi-Fi networks simultaneously inject MAC-layer packets into their protocol stacks and then transmit them on the shared channel. 
In this analysis we highlight (i)  the throughput of the secondary system and (ii) the interference generated to the primary Wi-Fi system when employing CoBeam, for different beamforming techniques and varying numbers of antennas at the secondary U-LTE transmitter.
Specifically, four transmission schemes are considered for the U-LTE transmission, i.e., (i) 
\textit{TX-1Ant}  (single-antenna omnidirectional),  (ii) \textit{ZF-2Ant} (Zero-Forcing Beamforming with two antennas),  (iii)  \textit{ZF-4Ant} (Zero-Forcing Beamforming with four antennas), and (iv) \textit{MRT-4Ant} (Maximum Ratio Transmission Beamforming with four antennas).
Then, we present an extensive experimental campaign measuring the aggregate throughput performance of the two spectrum sharing coexisting technologies for the \textit{ZF-4Ant} and \textit{MRT-4Ant} beamforming schemes while varying the locations of the secondary U-LTE receiver and the primary Wi-Fi users.

 \begin{figure}[t]
    \centering
    \includegraphics[width=.6\columnwidth]{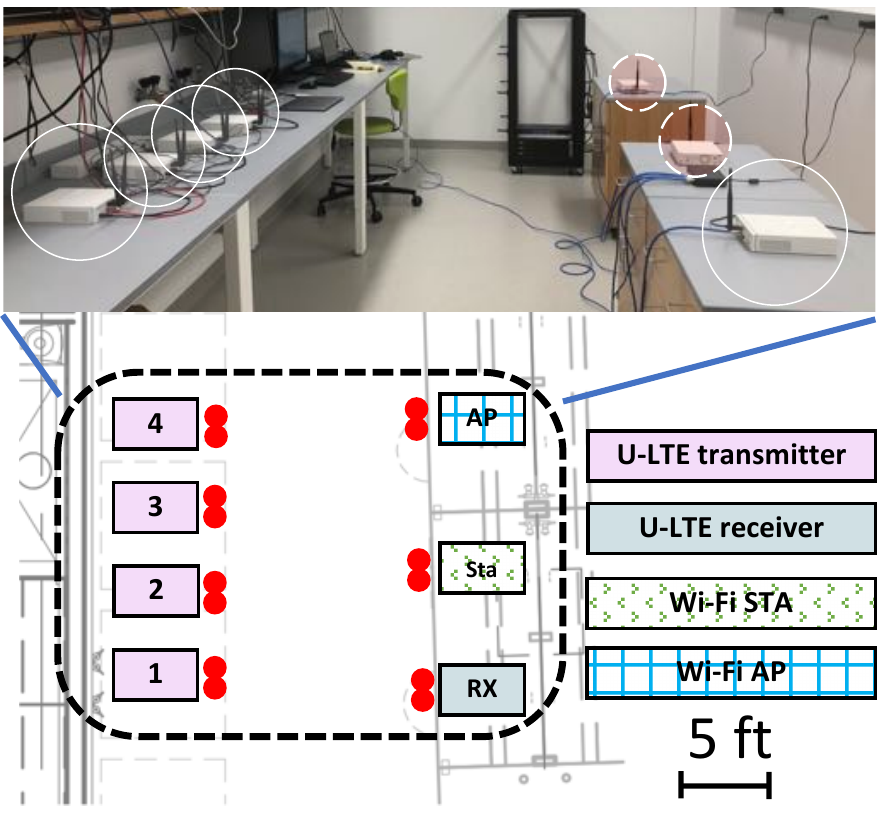}
    \caption{ \small Network deployment in Scenario 1.}
    \label{fig:bluprint1}
\end{figure}

\begin{figure}[t]
\centering
\includegraphics[height=.4\columnwidth , width=1\columnwidth]{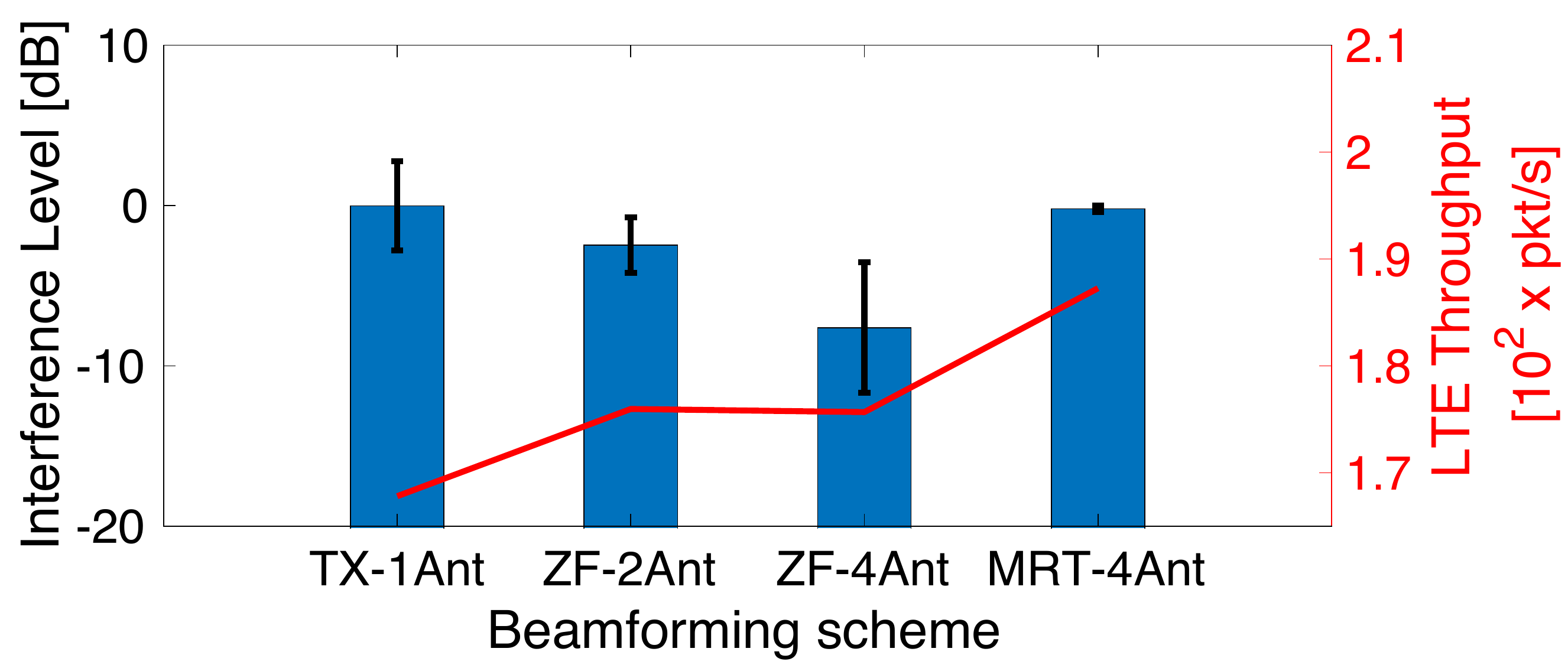}
\caption{ \small Throughput of U-LTE (curve) and corresponding caused interference levels (bars) in Scenario~1 for different beamforming schemes.}
\label{fig:received_power}
\vspace{-5mm}
\end{figure}

\noindent
\subsection{Experimental Results}
Figure~\ref{fig:bluprint1} illustrates a network scenario where the primary and secondary spectrum access technologies are co-located in a small room with significant mutual interference from simultaneous transmissions (referred to as Scenario 1). We first test the effectiveness of CoBeam in efficiently addressing the spectrum sharing challenges, by measuring the  interference generated at the Wi-Fi users when the U-LTE transmitter accesses the channel, as well as the corresponding throughput of the secondary networks, for four different beamforming schemes.
The average results of ten 1-minute long experiments are reported in Fig.~\ref{fig:received_power}. 
The performance comparison of the four beamforming schemes presented in Fig.~\ref{fig:received_power} shows how under \textit{MRT-4Ant} the U-LTE network achieves the highest throughput among the four tested schemes, obtaining a $23\%$ throughput gain with respect to single antenna omnidirectional \textit{TX-1Ant} ($215\;\mathrm{packets/s}$ vs $174\;\mathrm{packets/s}$). 
At the same time, since \textit{MRT-4Ant} aims at maximizing the throughput of the U-LTE network, the interference to the primary system can be as high as that with \textit{TX-1Ant} (i.e., no beamforming). 
Differently, the two ZF-based beamforming schemes achieve a better compromise between secondary system throughput and interference caused to Wi-Fi users. 
Specifically, under \textit{ZF-2Ant} and \textit{ZF-4Ant} the U-LTE network achieves approximately $12\%$ throughput gain with respect to the omnidirectional scheme ($192\;\mathrm{packets/s}$). 
When compared to \textit{TX-1Ant} and \textit{MRT-4Ant} scheme, \textit{ZF-2Ant} reduces the interference to the primary Wi-Fi network by $2\;\mathrm{dB}$ through ZF-based beamforming with two antennas; while \textit{ZF-4Ant} achieves up to $8\;\mathrm{dB}$ of interference reduction gain with respect to the mentioned schemes thanks to the greater degrees of freedom for interference nulling deriving from the ratio $N_t / \big(S_s + K_r \big)$ discussed in Section \ref{sec:wifilte}. 
In short, we experimentally verified how larger antennas/receivers ratios help to overcome the real-system CSI inaccuracies and improve the beamforming effectiveness.

\begin{figure}[t]
\centering
\includegraphics[width=.85\columnwidth
]{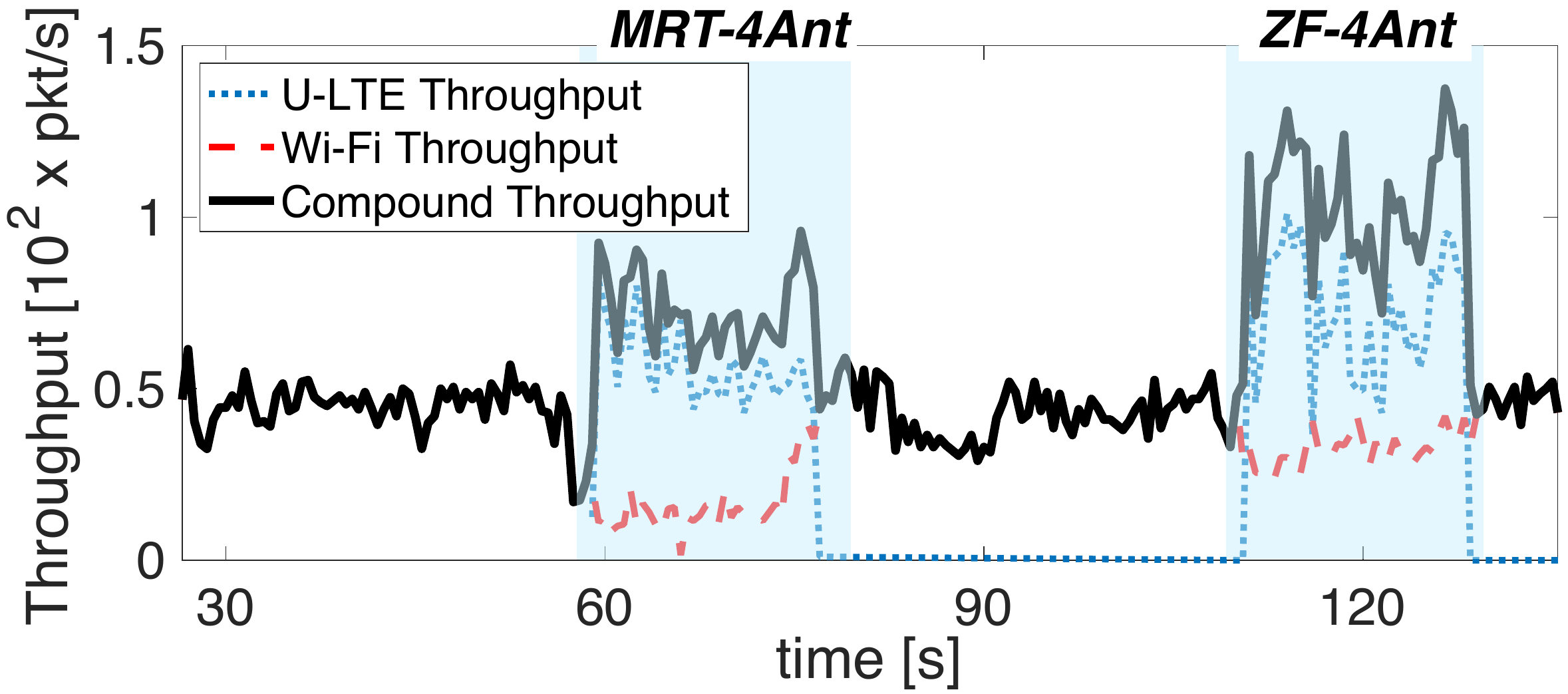}
\caption{ \small Single-run U-LTE and Wi-Fi throughput measures for different beamforming schemes over time in Scenario~1.}
\label{fig:1tx_vs_4zf}
\vspace{-0mm}
\end{figure}

Figures~\ref{fig:1tx_vs_4zf} provides a closer look at the interactions between the coexisting Wi-Fi and U-LTE networks on a single-run experiment where the Wi-Fi network generates traffic at an average rate of $100\;\mathrm{packets/s}$, while the U-LTE network accesses the channel twice, employing \textit{TX-1Ant} and \textit{ZF-4Ant} beamforming scheme, respectively,  $60\mathrm{s}$ and  $110\mathrm{s}$ after the beginning of the experiment. 
In Fig.~\ref{fig:1tx_vs_4zf}, we illustrate the measured instantaneous throughput achieved by the two coexisting networks for this experiment in Scenario~1 (Fig.~\ref{fig:scenario}). 
It can be seen starting from $60\mathrm{s}$ that the throughput of the Wi-Fi network gets significantly degraded by the U-LTE activity when no interference-nulling beamforming scheme is used (\textit{ZF-4Ant}), while the degradation is only marginal from second $110\mathrm{s}$ when \textit{ZF-4Ant} is employed.
When employing \textit{ZF-4Ant} at the secondary U-LTE, the aggregate primary and secondary network achieves up to $22\%$ throughput gain. 

\begin{figure}[t]
\centering
\includegraphics[width=.9\columnwidth]{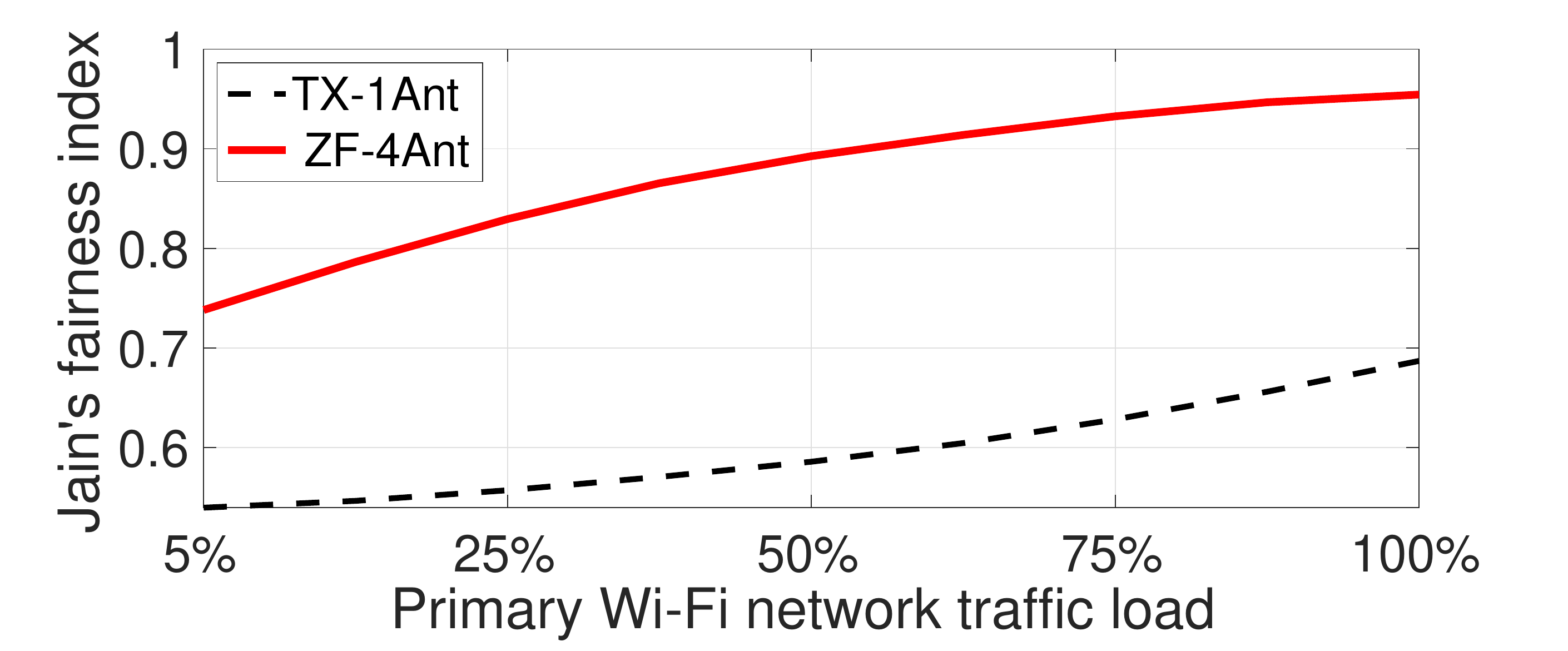}
\caption{ \small Network's Jain's fairness Index for different beamforming schemes in Scenario 1.}
\label{fig:fairness}
\vspace{-0mm}
\end{figure}

\noindent
\textit{Fairness:}
Here, we demonstrate that CoBeam is a spectrum sharing solution that is fair to unlicensed users. 
When employing \textit{ZF-4Ant} CoBeam, the average throughput is $75\;\mathrm{packet/s}$ for the primary Wi-Fi network and $230\;\mathrm{packet/s}$ for the secondary U-LTE network, with a resulting Jain's fairness index of $0.83$ when the primary system is operating at $~50\%$ of the load. 
Higher spectrum efficiency is obtained while guaranteeing lower interference to the primary system with respect to omnidirectional transmission, ensuring desirable fairness levels of the overall compound system.  
Additional results are presented in Fig.~\ref{fig:fairness} 
by considering different levels of Wi-Fi traffic load in Scenario~1. 
We observe that beamforming-based spectrum sharing always achieves better fairness, which ensures harmonious cross-technology coexistence in the same spectrum bands.

\textbf{Large scale experiments.}
In the section above, we showed how CoBeam scales with the number of antennas, and we analyzed the interference levels generated by the different beamforming schemes that we considered. Here, we compare the compound U-LTE and Wi-Fi network performances of \textit{ZF-4Ant} and \textit{MRT-4Ant} against single antenna  \textit{TX-1Ant} spectrum access scheme on larger scale experimental topologies.

\begin{figure}[t]
\centering
\includegraphics[, width=0.8\columnwidth]{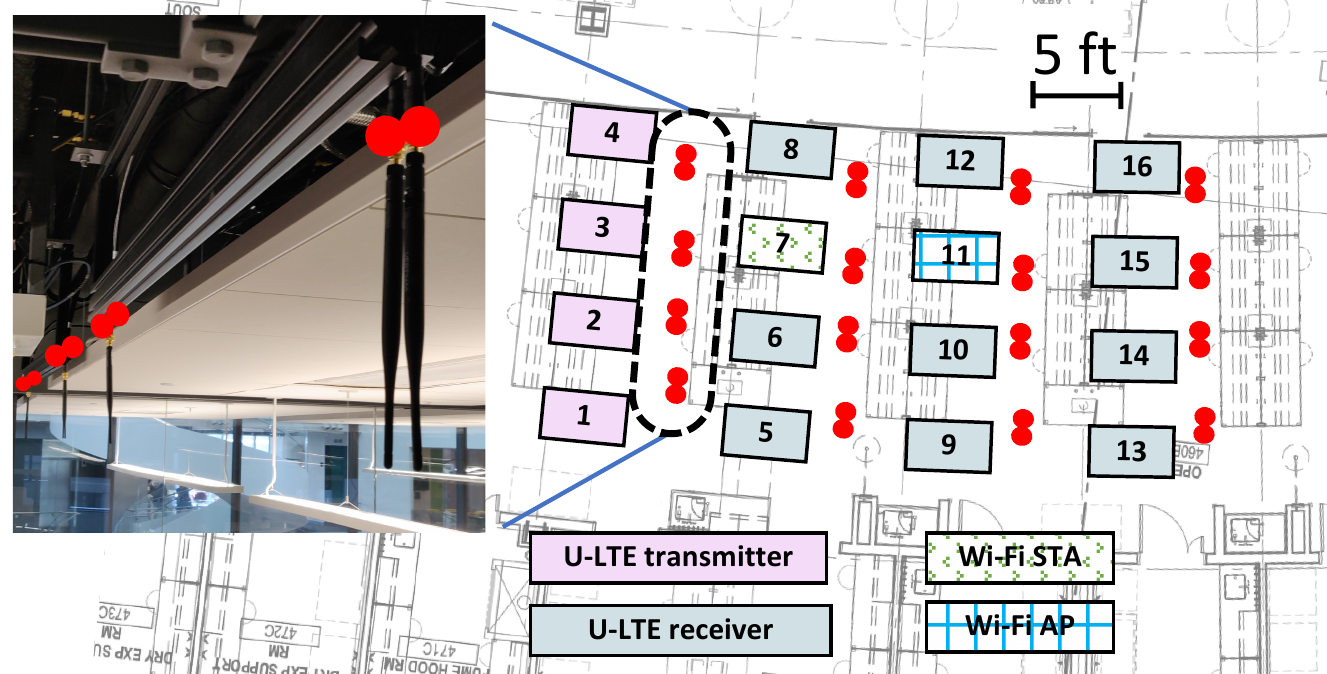}
\vspace{-0mm}
\caption{\small Network deployment in Scenario 2.}
\label{fig:scenario2}
\vspace{-0mm}
\end{figure}

In this series of experiments, we leveraged Arena \cite{BertizzoloMmnets19}, a large-scale Radio Frequency SDR-based indoor test grid.  
Arena features a mix of USRP N2210 and USRP X310 driving a set of antennas hanging off the ceiling of a typical office-like environment in a geometrical $8\times8$ grid layout, for a total of $64$ locations. 
In our experiments, we employed all the available USRP N210 on Arena, for a total of $16$ locations distributed in a $4\times4$ grid layout (see Fig. \ref{fig:scenario2}).
This testing environment represents well typical deployment scenarios of Wi-Fi/Cellular coexisting technologies, both in terms of scale and channel characteristics.

\begin{figure}[t]
\centering
\includegraphics[, width=0.9\columnwidth]{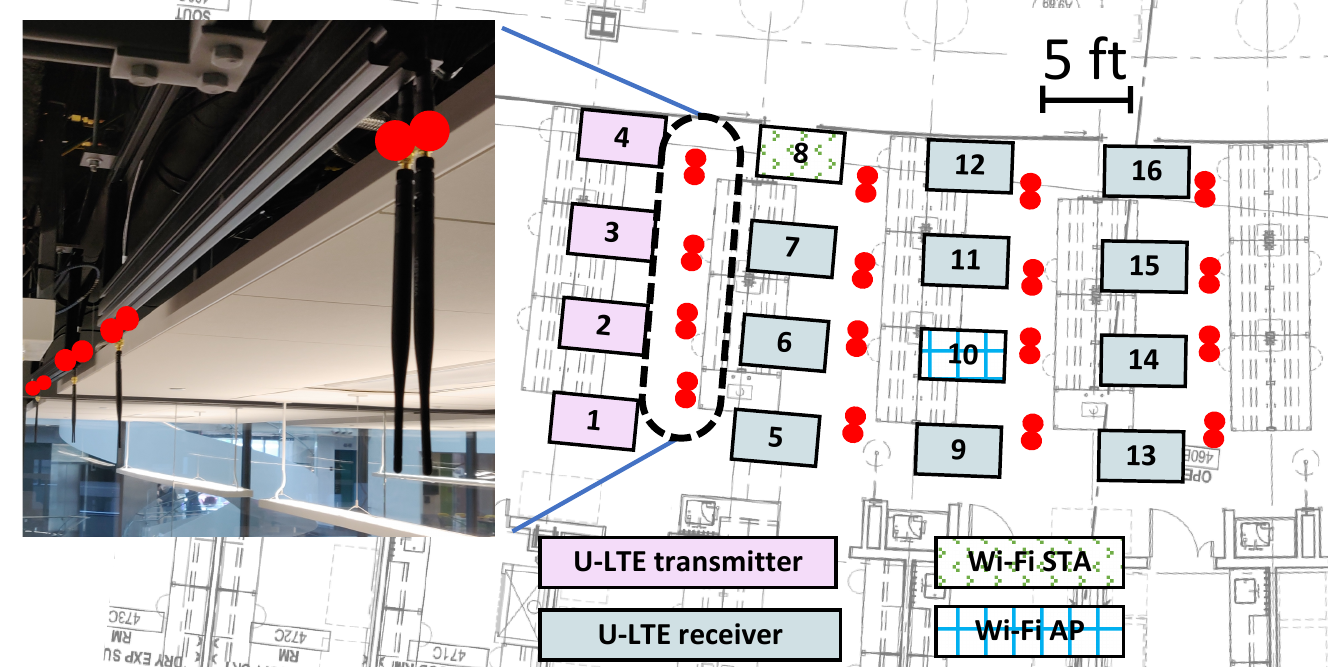}
\vspace{1mm}
\caption{\small Network deployment in Scenario 3.}
\label{fig:scenario3}
\vspace{2mm}
\end{figure}

We consider the two deployment scenarios shown in Fig.~\ref{fig:scenario2} and Fig.~\ref{fig:scenario3}, respectively. 
Scenarios 2 and 3 present a $4$-antenna secondary U-LTE transmitter located at the left end side of the testbed (SDRs $1-4$) and a primary Wi-Fi users pair, Wi-Fi Access Point (AP) and Wi-Fi Station (STA).
The Wi-Fi AP and the Wi-Fi STA are implemented at SDR $11$ and SDR $7$ in Scenario 2, and at SDR $10$ and SDR $8$ in and Scenario 3, respectively.
In both  scenarios, we evaluated the performance of the beamforming schemes  for all possible secondary U-LTE receiver locations across the testbed, as illustrated in Fig. \ref{fig:scenario2} and Fig. \ref{fig:scenario3}.
In the following experiments, we consider two coexisting primary and secondary networks, for two different primary network traffic activity patterns, low traffic load ($5\%$ of full bandwidth) and high load ($100\% $ of full bandwidth), respectively. 
We evaluate the performances of the compound network system implementing \textit{MRT-4Ant} beamforming at the U-LTE transmitter when the Wi-Fi traffic activity is low and adopting \textit{ZF-4Ant} at U-LTE when the Wi-Fi traffic load is high, respectively.

\begin{figure}[t]
\centering
\includegraphics[, width=1\columnwidth]{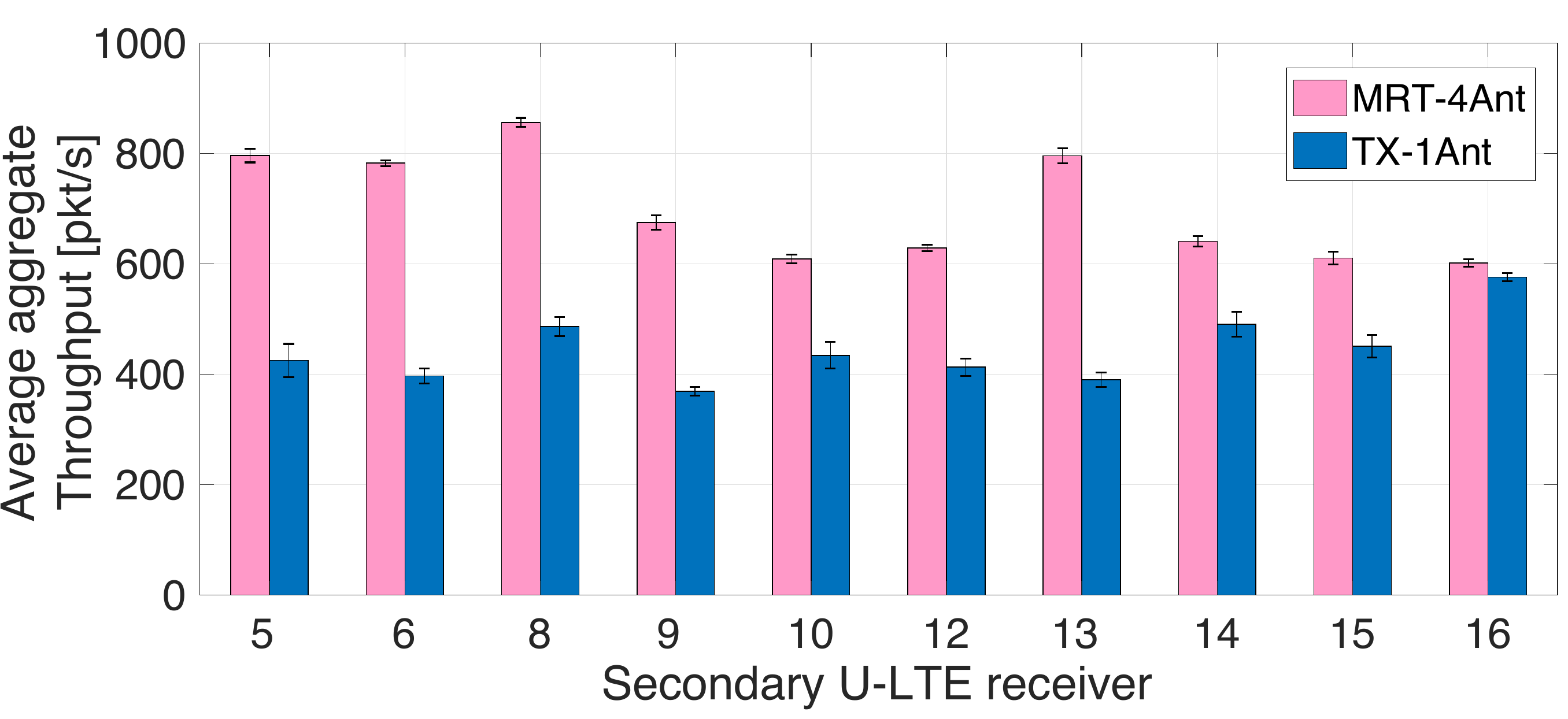}
\vspace{-5mm}
\caption{\small MRT beamforming and omnidirectional transmission performances comparison for Scenario 2.}
\label{fig:mrtscenario2}
\vspace{-5mm}
\end{figure}

\noindent
\textit{Maximum Ratio Transmission Beamforming:}
In Scenario 2 (see Fig. \ref{fig:scenario2}) we evaluate the performance of \textit{MRT-4Ant} beamforming scheme compared with \textit{TX-1Ant} for $10$ different secondary U-LTE receiver locations for low primary system traffic load. 
We present the average aggregate performance of the compound Wi-Fi and U-LTE network for \textit{MRT-4Ant} and \textit{TX-1Ant} over ten 1-minute long experiments in Fig. \ref{fig:mrtscenario2}, which presents the overall network throughput accounting for both primary and secondary transmissions for $10$ different secondary receiver locations.
Under \textit{MRT-4Ant}, the secondary receivers benefit from favorable distortion-free effective channel conditions, as per  (\ref{eq:mrt_ffective_channel}) in Section \ref{sec:problem}, and higher SINR, as in  (\ref{eq:snr_mrt}) in Section \ref{sec:problem}. 
This results in improved throughput performance compared to single-antenna omnidirectional transmission schemes.
When employing  \textit{TX-1Ant}, the aggregate Wi-Fi and U-LTE network achieves up to $400$ more packets per second with respect to  \textit{MRT-4Ant}, with an aggregate average performance gain of $57\%$ over $10$ different topologies.

\noindent
\textit{Zero-Forcing Beamforming:}
When the primary system traffic load is high, MRT beamforming fails to achieve good spectrum utilization performance and maintain high network fairness.
In such situations, CoBeam has been designed to implement Zero-Forcing Beamforming-based transmissions.
We evaluated the performance of Zero-Forcing beamforming against single-antenna omnidirectional transmission in two deployment scenarios, namely Scenarios 2 and 3 (Fig. \ref{fig:scenario2} and Fig. \ref{fig:scenario3}), by varying the location of the secondary U-LTE receiver, for a total of $18$ different network topologies.
Average aggregate network throughput for
over ten 1-minute long experiments are presented in Fig. \ref{fig:zfscenario2} and Fig. \ref{fig:zfscenario3}, respectively.
In both scenarios, and for all the considered topologies, CoBeam is proven to achieve higher aggregate network performance with peak throughput gains of $508$~pkt/s in Scenario 2 and $267$~pkt/s in Scenario 3; and with a significant overall average performance improvement of $169\%$.

\subsection{Final Remarks}

\begin{figure}[t]
\centering
\includegraphics[, width=1\columnwidth]{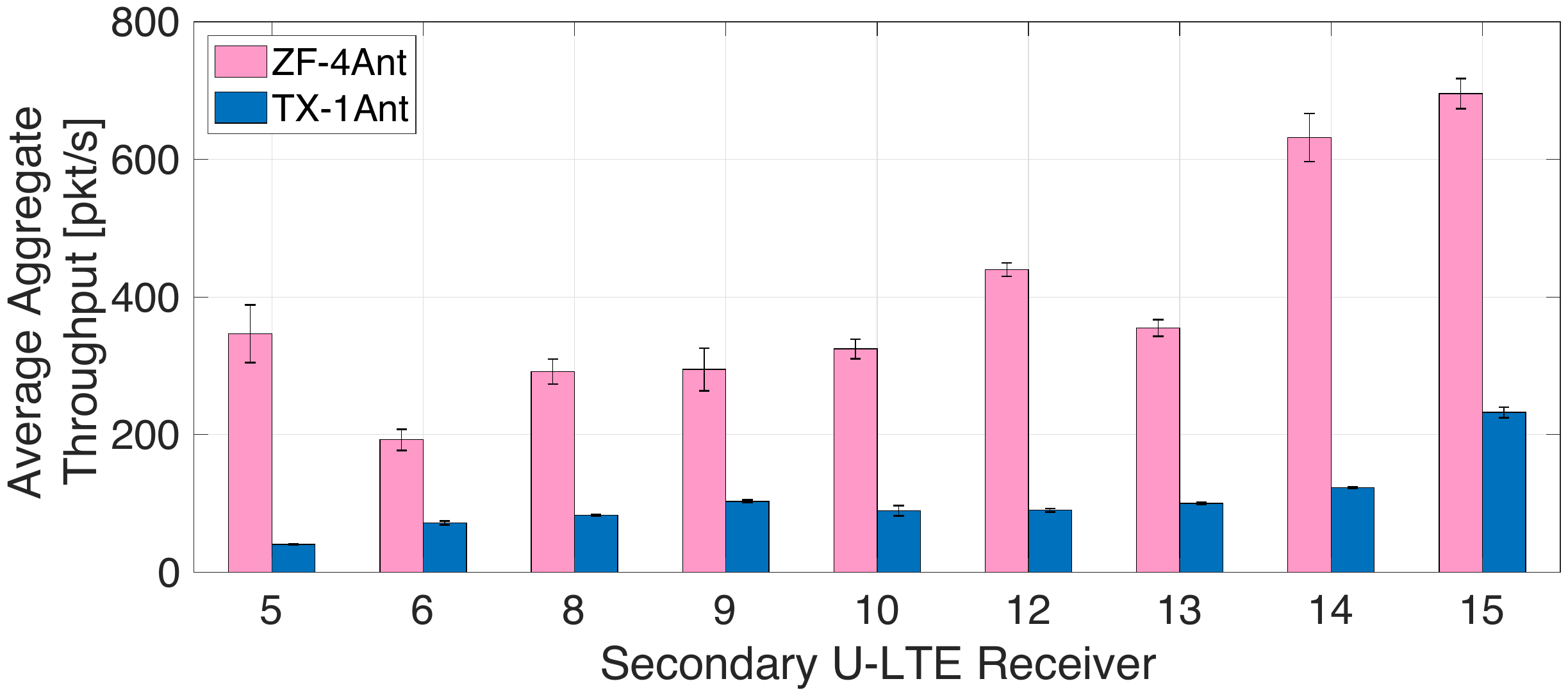}
\vspace{-5mm}
\caption{\small ZF beamforming and omnidirectional transmission performances comparison for Scenario 2.}
\label{fig:zfscenario2}
\vspace{-0mm}
\end{figure}

\begin{figure}[t]
\centering
\includegraphics[, width=1\columnwidth]{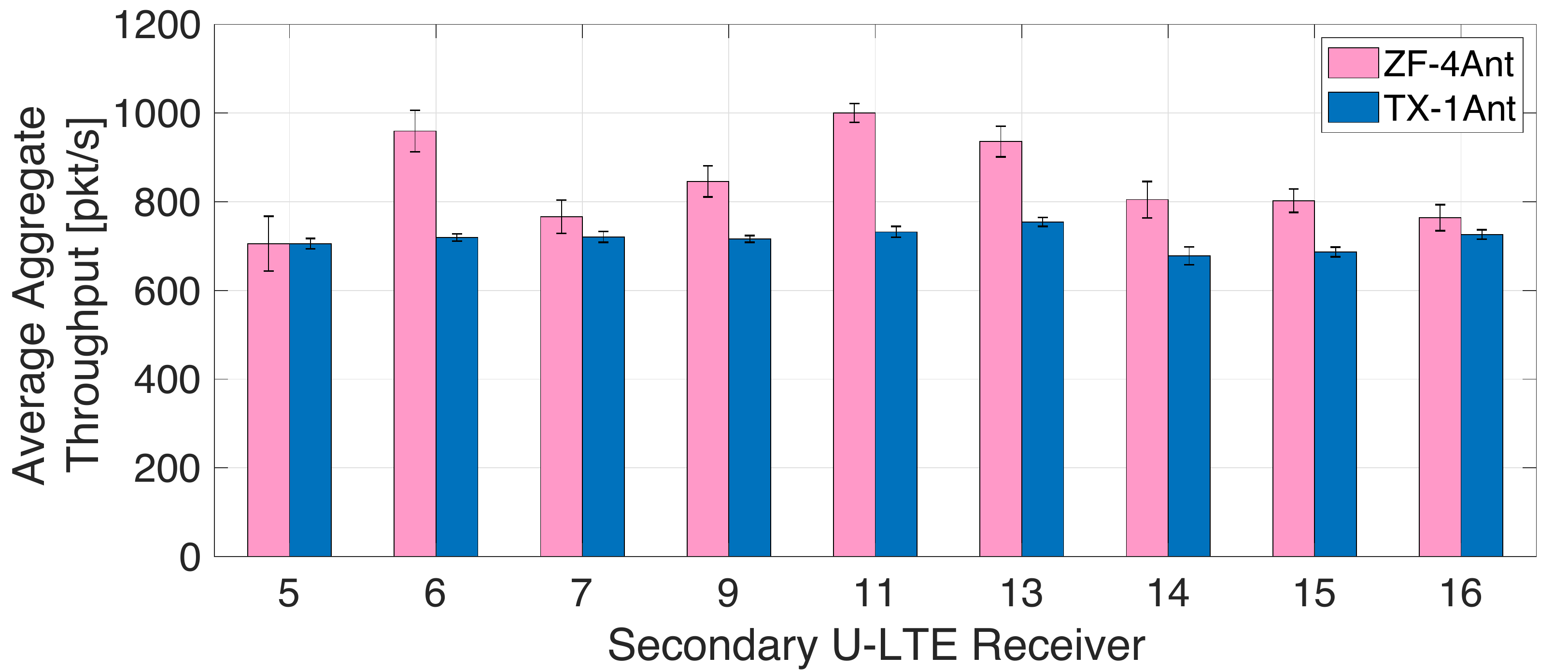}
\vspace{-5mm}
\caption{\small ZF beamforming and omnidirectional transmission performances comparison for Scenario 3.}
\label{fig:zfscenario3}
\vspace{-3mm}
\end{figure}

\textit{Mobility:} Results presented in Figs. \ref{fig:mrtscenario2},  \ref{fig:zfscenario2}, and  \ref{fig:zfscenario3} can be considered to be representative of mobility emulation. The reported results capture the performance of CoBeam when rapidly switching the secondary U-LTE receiver. 
CoBeam continuously computes new beamforming coefficients to match stringent channel coherence timing (roughly $100\:\mathrm{ms}$ for sub-$6\:\mathrm{GHz}$ bands), while human/pedestrian mobility is on a larger time scale, in the order of seconds. 

\textit{Hidden Terminals and Interference:}
In this work, we focused on the problem of secondary transmitter spectrum access in populated primary system bands. 
In designing CoBeam, we relied on the assumption of homogeneous noise and interference levels across the deployment area, and single-antenna receivers. 
Network configurations involving hidden terminals and heterogeneous interference levels might require more advanced precoding schemes and multiple-antenna interference cancellation techniques at the receivers.

\textit{Complexity, Delay, and Scalability:}
CoBeam's SDR-based prototype was developed to provide a proof of concept as well as realistic performance evaluation of cognitive-beamforming-based spectrum sharing schemes. 
It is worth mentioning that current SDR-based prototypes suffer from inherent radio-to-host latency and processing time jitter typical of OS-based baseband processing, which results in limited scalability due to high computational capabilities demands. 
In our future work, we will consider software implementations based on real-time operating systems, as well as low-latency hardware-based processing solutions based on FPGAs.

\section{Conclusions} \label{sec:conclusion}
We presented the basic building blocks of CoBeam, a new beamforming-based cognitive framework enabling \mbox{spectrally-efficient} channel coexistence for heterogeneous wireless networks with different medium access technologies on the same spectrum bands. 
We discussed the challenges
to enable simultaneous channel access with no cross-technology communication, thus overcoming traditional cross-technology signaling based methods to achieve harmonious coexistence in 5G wireless networks with shared spectrum. 
We also discussed a prototype of CoBeam for LTE and Wi-Fi coexistence in unlicensed bands, and proved its effectiveness and flexibility through extensive experimental results. 
We reported an average of $169\%$ network throughput gain and $8\;\mathrm{dB}$ interference power reduction for high Wi-Fi traffic loads, and an average of $57\%$ network throughput gain for low Wi-Fi traffic loads. 

Future research directions will include enhancing the cognitive sensing engine with additional cognitive approaches such as blind channel estimation, and extending the prototype to FPGA implementation and to a wider range of coexisting technologies.

\clearpage
\bibliographystyle{ieeetr}
\bibliography{NOS,Mobihoc2010,WiFiOpt3,WiFiOpt2}

\end{document}